\documentclass[%
 reprint,
 superscriptaddress,
 amsmath,amssymb,
 aps,
]{revtex4-1}

\usepackage{graphicx}
\usepackage{dcolumn}
\usepackage{multirow}
\usepackage{bm}
\usepackage{xcolor}
\usepackage{ulem}

\begin{document}


\title{A sign of three-nucleon short-range correlation from
an analysis of nuclear mass and short-range correlation probability}

\author{Na-Na Ma}
\email{mann@lzu.edu.cn}
\affiliation{School of Nuclear Science and Technology, Lanzhou University, Lanzhou 730000, China}

\author{Rong Wang}
\email{Corresponding author: rwang@impcas.ac.cn}
\affiliation{Institute of Modern Physics, Chinese Academy of Sciences, Lanzhou 730000, China}
\affiliation{School of Nuclear Science and Technology, University of Chinese Academy of Sciences, Beijing 100049, China}


\date{\today}

\begin{abstract}
Three-nucleon short-range correlation ($3N$ SRC) represents a rare
and intriguing part of the nuclear dynamics at short distance,
beyond the two-nucleon short-range correlation ($2N$ SRC).
To search its existence is a hot topic in the ongoing and future
high-energy nuclear experiments and the developments of nuclear theory.
In this study, we found a positive sign of $3N$ SRC in nuclei,
by analyzing the correlation between the per-nucleon nuclear mass
and the probability of a nucleon in $2N$ SRC state,
with the current experimental measurements of $^2$H, $^3$He, $^4$He, $^9$Be,
$^{12}$C, $^{27}$Al, $^{56}$Fe, Cu, $^{197}$Au and $^{208}$Pb
from SLAC, CLAS, and JLab Hall C collaborations.
The effective masses of the nucleons in $2N$ SRC and
$3N$ SRC are also extracted from the analysis,
which provide some references for the nuclear medium effect study.
The probability of $3N$ SRC is much smaller than that
of $2N$ SRC, thus requiring high-luminosity experiments
to confirm its existence.
\end{abstract}

\maketitle


\section{Introduction}\label{sec:intro}

A breakthrough in high-energy electron/proton-scattering experiments
off the nuclear targets initiated
the study of the nucleon-nucleon short-range correlation
\cite{Frankfurt:1993sp,Aclander:1999fd,Tang:2002ww,Piasetzky:2006ai,CLAS:2005ola,Subedi:2008zz,Hen:2014nza,CLAS:2018yvt,CLAS:2019vsb},
which is a novel phenomenon driven by nuclear dynamics range
from medium distance to short distance
\cite{Frankfurt:1988nt,Sargsian:2005ru,Vanhalst:2011es,Weiss:2016obx,Schiavilla:2006xx,Alvioli:2007zz,Neff:2015xda}.
A surge of interests in the SRC research,
with a significant amount of experiments and theoretical progresses,
provided the criteria for the identification of SRC,
and gradually revealed the underlying physical mechanism of SRC
and the significance of SRC in the frontiers of nuclear physics and astrophysics.
Measurements from both electron-nucleus ($eA$) and proton-nucleus ($pA$) experiments
finally lead to some consensuses regarding nucleon-nucleon SRC:
the nucleon inside SRC pair has a large relative momentum compared
with the Fermi momentum ($k_F$) of the system
while the SRC pair as a whole has a small center-of-mass momentum,
and under the interplay of tensor force and short-range repulsive force,
the SRC pair has two dominant features of locality and high momentum
\cite{Hen:2016kwk,Arrington:2011xs,Fomin:2017ydn,Arrington:2022sov}.

Multi-nucleon short-range correlations represent the most intriguing part
of the nuclear dynamic beyond the shell model
and they are closely related to the strong interaction physics
at the quark level as well as the relevant high-density dynamics in neutron star
\cite{Subedi:2008zz,Frankfurt:2008zv}.
(i) SRC occurs in short-distance region,
hence it is closely related to the phenomena governed by quantum chromodynamics (QCD),
including chiral symmetry, quark interchanges between nucleons,
and quark-gluon degrees of freedom, to name just a few examples.
(ii) An ingenious analogy between the waltz and the nucleon-nucleon SRC \cite{Hen:2014nza}
vividly explains a universal nature of the dominant $np$ SRC pairs
in both symmetric and asymmetric nuclei.
This is vital for understanding the dynamics in the interior
of neutron stars with a small fraction of protons.
(iii) A remarkable linear correlation between the strength of
the European Muon Collaboration (EMC) effect
and the nucleon-nucleon SRC probability links these two seemingly disconnected phenomena,
and implies that the nucleon-nucleon SRC induces
a substantial change of the structure of bound nucleon
thus yields the nuclear EMC effect
\cite{CLAS:2019vsb,Weinstein:2010rt,Hen:2012fm,Arrington:2012ax,Chen:2016bde,Wang:2022kwg,Ma:2023tsi}.
In view of these aspects, it has been an area of continuing interests
to investigate the details of multi-nucleon short-range correlations.

Up to date, the properties and details of 2N SRC have been
extensively investigated, while the existence of multi-nucleon
SRC of more two nucleons is still not clear \cite{HallA:2017ivm,Sargsian:2019joj,Day:2018nja}.
Beyond the 2N SRC, the main subject of this study is 3N SRC.
First of all, it is an interesting topic
whether there is a relationship between the 3N SRC
and the three-body nuclear force \cite{Sargsian:2005ru,Sargsian:2019joj,Day:2018nja}.
Second, from our previous studies, we found that
the 2N SRC only is not enough to reproduce
the observed nuclear EMC effect within both
a $x$-rescaling model and a nucleon-swelling model \cite{Wang:2022kwg,Ma:2023tsi}.
Besides the dominant 2N SRC,
3N SRC could be another origin of the EMC effect.
Third, 3N SRC is an important intermediate state in
forming the four-nucleon SRC and other types of nucleon clusters.
Therefore, it is imperative to study the $3N$ SRC,
so as to unveil the diverse and new micro structures
inside nucleus.

The $3N$ SRC is a type of short-distance configuration of nucleons
with small inter-distances ($\leq 1.2$ fm),
which has the very similar characteristics of the two-nucleon SRC \cite{Hen:2016kwk,Fomin:2017ydn}.
The nucleon momentum in $3N$ SRC significantly exceeds $k_F$,
but in this case of $3N$ SRC, the high-momentum nucleon is balanced
by the other two correlated nucleons with momenta around $k_F$.
The c.m. momentum of the 3N SRC pair $p_{cm}$ is argued
to be less than or equal to $k_F$ \cite{Sargsian:2019joj,Day:2018nja}.
On the experimental side, the appearance of a plateau of cross-section ratio
in the region of the Bjorken variable $x_B\geq 2$ is the most direct
signal for the existence of 3N SRC \cite{HallA:2017ivm}.
The pioneering experimental result by CLAS collaboration
more than one decade ago displayed the second plateau
after the $2N$ SRC plateau with large uncertainties as well,
which hints the existence of $3N$ SRC \cite{CLAS:2005ola}.
However the subsequent experiment by JLab Hall C collaboration
reported that the $3N$ SRC plateau disappears confronting
the more precise experimental data.
The recent analysis of the CLAS data found that
this previous observed $3N$ SRC plateau is an effect
from the bin migrations \cite{Higinbotham:2014xna}.
It is worth mentioning that the E02-019 experiment at JLab Hall C
display no $3N$ SRC plateau at all,
with the assistance of all possible experimental technologies.
At present, there are more experiments planed to intensively
verify the existence of $3N$ SRC, and parallel with the future experiments,
exploring more new methods to pin down the $3N$
short-range correlations is also a feasible direction in the field.

In this paper, we try to find an evidence or a hint
of $3N$ SRC from an in-depth analysis of the nuclear
mass in terms of the 2N-SRC probability.
In Sec. \ref{sec:SRC-probability}, we introduce
the definitions of 2N-SRC and 3N-SRC probabilities
that will used in this analysis, and the relation
between the 2N-SRC probability and the 3N-SRC probability.
The decomposition of nuclear mass in terms of
mean-field nucleons and multi-nucleon short-range
correlations is given in Sec. \ref{sec:nuclear-mass}.
The correlation between the nuclear mass and
the SRC probability is analyzed with and without
the 3N SRC, which is shown in Sec. \ref{sec:correlation-analysis}.
Finally, some discussions and a summary of our analysis
are present in Sec. \ref{sec:discussions-summary}.

\section{Probabilities of 2-nucleon and 3-nucleon short-range correlations}\label{sec:SRC-probability}

\begin{figure*}[htbp]
\begin{center}
\includegraphics[width=0.92\textwidth]{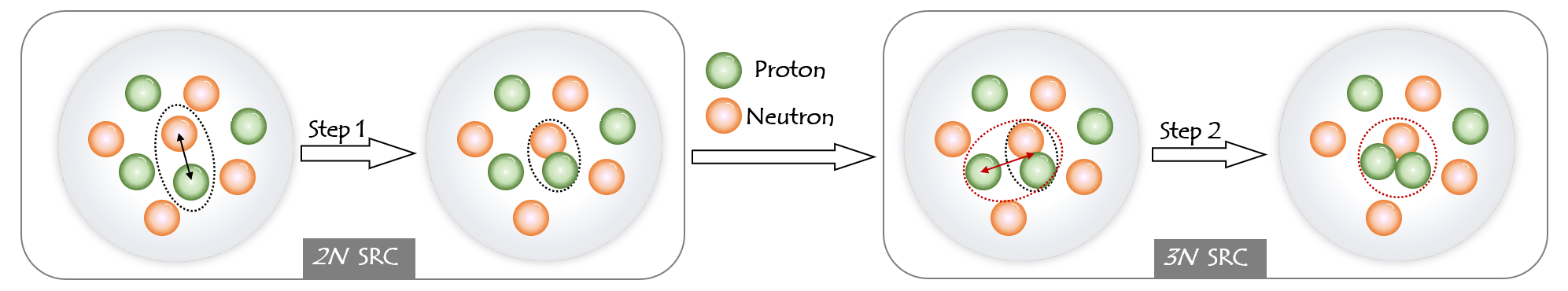}
\caption{The $3N$ SRC is formed in two steps. In the first step,
two nucleons forms a $2N$ SRC. In the second step, a nucleon and
the $2N$ SRC attract each other and combine into a $3N$ SRC. }
\label{3NSRC-formation-process}
\end{center}
\end{figure*}

In this study, the 2N SRC probability is defined as
the probability of a nucleon being in a 2N SRC pair,
which is written as,
\begin{equation}
\begin{aligned}
  P_{\rm 2N} \equiv \frac{2\times n_{\rm 2NSRC}} {A},
\end{aligned}
\label{P2N-def}
\end{equation}
where $n_{\rm 2NSRC}$ is the number 2N SRC pairs
and $A$ is the total number of nucleons.
Similarly, the 3N SRC probability is defined as,
\begin{equation}
\begin{aligned}
  P_{\rm 3N} \equiv \frac{3\times n_{\rm 3NSRC}} {A},
\end{aligned}
\label{P3N-def}
\end{equation}
in which $n_{\rm 3NSRC}$ is the number of 3N SRC clusters.
With decades of efforts by experimentalists and theorists,
the number of 2N SRC pairs in nucleus is already quantified
\cite{Frankfurt:1993sp,CLAS:2005ola,Fomin:2011ng,CLAS:2019vsb}.
Hence, the 2N SRC probability also can be estimated.
However, there is few information about the 3N SRC in experiments,
and its existence also need to be confirmed.
It is not easy to have a reliable estimation of
the 3N SRC probability.

In this work, we make an estimation of 3N SRC probability
based on a simple assumption that the 3N SRC cluster is formed
with sequential processes, and it parameterized as $P_{\rm 3N}=\kappa P^2_{\rm 2N}$
in which $\kappa$ is a free parameter.
Fig. \ref{3NSRC-formation-process} schematically shows how the 3N SRC configuration
is produced with two sequential processes of cohesion.
The first process is that two nucleons attract each other and form into
the close-proximity configuration as a 2N SRC pair.
If luckily, within the short lifetime of the 2N SRC state,
another nucleon is attached to the 2N SRC and generate a 3N SRC cluster.
This is the second process for the 3N SRC formation.
The probabilities of the first and the second processes are similar,
for the similar intermediate attraction force is needed
for both the first and the second processes.
Therefore, the probability for the second process to happen is $\kappa P^2_{\rm 2N}$,
with the parameter $\kappa$ is probably near 1.
Then the probability of a nucleon in a 3N SRC is given by,
\begin{equation}
\begin{aligned}
  P_{3N} = P_{\rm 2N} \times \kappa P_{\rm 2N} = \kappa P_{\rm 2N}^2.
\end{aligned}
\label{P3N-formula}
\end{equation}
Combing Eqs. (\ref{P2N-def}) and (\ref{P3N-formula}),
finally we have an estimate of $P_{3N}$, which is written as,
\begin{equation}
\begin{aligned}
  P_{3N} = \kappa \left( \frac{2 n_{\rm 2NSRC}}{A} \right)^2,
\end{aligned}
\label{P3N-formula-final}
\end{equation}
in which $\kappa$ is a free parameter probably around 1.

Note that Fig. \ref{3NSRC-formation-process} just shows a simple picture of cluster formation processes.
As the nucleus is complex many-body system,
rigorous theoretical calculation on the probability of
the cohesion process is very challenging,
yet looking forward in the future.
For multi-nucleon short-range correlations,
the quark-exchange and the gluon-exchange processes also should be
considered for the short-distance interactions.
Nevertheless, according to the simple physical mechanism
shown in Fig. \ref{3NSRC-formation-process}, the probability of short-range correlations
of more nucleons goes down quickly.

\section{Nuclear mass from short-range correlations}\label{sec:nuclear-mass}

A simple global property of a nucleus is the mass,
which is closely related to the nuclear binding or the mass deficit.
Although the nucleus is a complex quantum system of
complicated nucleon motions and diversified micro-structures,
the sophisticated nuclear structure should be reflected
in the nuclear mass. More importantly,
the masses of most nuclei are already precisely measured in experiments.
In this work, we try to analyze the nuclear mass at length
in terms of the micro-structures.

Suppose in a nucleus, the nucleons can be classified into
three categorises: the mean-field nucleon, the nucleon in 2N SRC,
and the nucleon in 3N SRC.
Then the nuclear mass can be decomposed as,
\begin{equation}
\begin{aligned}
  m_{\rm A} = &(A - A P_{\rm 2N} - A P_{\rm 3N})m_{\rm \bar{N}} \\
        &+ A P_{\rm 2N}m_{\rm \bar{N}}^{\rm 2N} + AP_{\rm 3N}m_{\rm \bar{N}}^{\rm 3N},
\end{aligned}
\label{nuclear-mass-decomposition}
\end{equation}
where $m_{\rm \bar{N}}$, $m_{\rm \bar{N}}^{\rm 2N}$, and $m_{\rm \bar{N}}^{\rm 3N}$
denote respectively the average mean-field nucleon mass,
the average nucleon mass in 2N SRC, and the average nucleon mass in 3N SRC.
If we remove the third term on the right side in Eq. (\ref{nuclear-mass-decomposition}),
then we just a nuclear mass decomposition without 3N SRC clusters.
By using this mass decomposition formula,
we assume that the properties of 2N SRC or 3N SRC are universal
among different nuclei. Actually, the approximate universality of 2N SRC
has been tested and predicted in experiments \cite{CLAS:2005ola,Fomin:2011ng,CLAS:2019vsb}
and theoretical calculations \cite{Feldmeier:2011qy,Alvioli:2016wwp}.
Therefore the proposed nuclear mass decomposition is a good approximation
of the nuclear mass in terms of some special micro-structures.
For the nuclear mass per nucleon,
rearranging Eq. (\ref{nuclear-mass-decomposition}), thus we have,
\begin{equation}
\begin{aligned}
  \frac{m_{\rm A}}{A} = m_{\rm \bar{N}} + (m_{\rm \bar{N}}^{\rm 2N}-m_{\rm \bar{N}}) P_{\rm 2N}
          + (m_{\rm \bar{N}}^{\rm 3N}-m_{\rm \bar{N}}) P_{\rm 3N}.
\end{aligned}
\label{per-nucleon-mass}
\end{equation}
Combing Eqs. (\ref{P3N-formula}) and (\ref{per-nucleon-mass}),
finally we have the per-nucleon nuclear mass
in terms of the probability of 2N SRC, which is written as,
\begin{equation}
\begin{aligned}
  \frac{m_{\rm A}}{A} = m_{\rm \bar{N}} + (m_{\rm \bar{N}}^{\rm 2N}-m_{\rm \bar{N}}) P_{\rm 2N}
          + (m_{\rm \bar{N}}^{\rm 3N}-m_{\rm \bar{N}}) \kappa P_{\rm 2N}^2.
\end{aligned}
\label{per-nucleon-mass-final}
\end{equation}

From Eq. (\ref{per-nucleon-mass-final}), one sees that
the per-nucleon nuclear mass is a quadratic function
of the 2N SRC probability, $a P_{\rm 2N}^2 + b P_{\rm 2N} + c$,
so long as there are 3N SRCs exist in the nucleus
and the 3N SRC is generated from the combination
of a nucleon and a 2N SRC pair.
At the same time, if there is no short-range correlation
of more than two nucleons, then the nuclear mass is just a
linear function of the 2N SRC probability, as $b P_{\rm 2N} + c$.
Therefore, by analyzing the correlation between the nuclear mass
and the probability of 2N SRC,
we could have some indications of multi-nucleon SRCs
beyond the 2N SRC.
If the nuclear mass is not linearly correlated
with the 2N SRC probability, then
there probably are some short-range correlations of more than two nucleons.

\section{Correlation between nuclear mass and short-range correlation probability}\label{sec:correlation-analysis}

\begin{table*}[h]
\caption{The experimental data of SRC scaling factor $a_2$ from
         SLAC, CLAS and JLab Hall C collaborations,
         and the resulting average value, for various nuclei. }
\renewcommand\arraystretch{1.25}
  \begin{tabular}{cccccc}
    \hline\hline
       Nucleus  & SLAC\cite{Frankfurt:1993sp} & CLAS06\cite{CLAS:2005ola} & CLAS19\cite{CLAS:2019vsb} & Hall C\cite{Fomin:2011ng} & Average \\
    \hline
       $^3$He   & 1.7$\pm$0.3 & 1.97$\pm$0.10 &                & 2.13$\pm$0.04 & 2.10$\pm$0.04  \\
       $^4$He   & 3.3$\pm$0.5 & 3.80$\pm$0.34 &                & 3.60$\pm$0.10 & 3.60$\pm$0.10  \\
       $^9$Be   &             &               &                & 3.91$\pm$0.12 & 3.91$\pm$0.12  \\
     $^{12}$C   & 5.0$\pm$0.5 & 4.75$\pm$0.41 & 4.49$\pm$0.17  & 4.75$\pm$0.16 & 4.65$\pm$0.11  \\
    $^{27}$Al   & 5.3$\pm$0.6 &               & 4.83$\pm$0.18  &               & 4.87$\pm$0.18  \\
    $^{56}$Fe   & 5.2$\pm$0.9 & 5.58$\pm$0.45 & 4.80$\pm$0.22  &               & 4.96$\pm$0.20  \\
           Cu   &             &               &                & 5.21$\pm$0.20 & 5.21$\pm$0.20  \\
    $^{197}$Au  & 4.8$\pm$0.7 &               &                & 5.16$\pm$0.22 & 5.13$\pm$0.21  \\
    $^{208}$Pb  &             &               & 4.84$\pm$0.20  &               & 4.84$\pm$0.20  \\
    \hline\hline
  \end{tabular}
\label{tab:a2-values}
\end{table*}

\begin{table*}[h]
\caption{The averaged value of nucleon-nucleon SRC scaling factor $a_2$,
        the probabilities of a nucleon in 2N SRC from two different models (I and II),
        the nuclear mass per nucleon, for various nuclei. }
\renewcommand\arraystretch{1.25}
\begin{center}
  \begin{tabular}{cccccccccccc}
    \hline\hline
      Nucleus  &  $^2$H  &  $^3$He  &  $^4$He  &  $^9$Be  &  $^{12}$C  & $^{27}$Al & $^{56}$Fe &  Cu  & $^{197}$Au & $^{208}$Pb \\
    \hline
      $a_2$    &   1     & 2.10$\pm$0.04 & 3.60$\pm$0.10 & 3.91$\pm$0.12 & 4.65$\pm$0.11 & 4.87$\pm$0.18 & 4.96$\pm$0.20 & 5.21$\pm$0.20 & 5.13$\pm$0.21 & 4.84$\pm$0.20  \\
      $P_{\rm 2N}^{\rm I}$ & 0.021(5) & 0.044(1) & 0.076(2) & 0.082(3) & 0.097(2) & 0.102(4) & 0.104(4)  & 0.109(4) & 0.108(4) & 0.102(4) \\
      $P_{\rm 2N}^{\rm II}$ & 0.041(8) & 0.086(2) & 0.148(4) & 0.160(5) & 0.191(5) & 0.200(7) & 0.203(8)  & 0.214(8) & 0.210(9) & 0.198(8) \\
      $\frac{m_{\rm A}}{A}$ [MeV] & 937.81 & 936.13 & 931.85 & 932.53 & 931.24 & 930.61 & 930.18  & 930.23 & 931.13 & 931.19  \\
    \hline\hline
  \end{tabular}
\end{center}
\label{tab:P2N-and-nuclear-mass}
\end{table*}

The 2N SRC probability in a nucleus is closely related to
the number of 2N SRC pairs in the nucleus,
according to the definition in Eq. (\ref{P2N-def}).
And the number of 2N SRC pairs in a nucleus is proportional
to the SRC scaling factor $a_2$ ($a_2$ is defined as the
quasielastic cross-section ratio between a heavy nucleus
and the deuteron, in the 2N SRC kinematical region).
Thanks to the developments of high-intensity and high-energy
scattering experiments off the nuclear targets,
the SRC scaling factor $a_2$ of many nuclei have been measured
by SLAC \cite{Frankfurt:1993sp}, CLAS \cite{CLAS:2005ola,CLAS:2019vsb}
and JLab Hall C collaborations \cite{Fomin:2011ng},
which are listed in Table \ref{tab:a2-values}.
The combined mean values are provided in the table as well.
With the measured value of $a_2$, the number of 2N SRC pairs
in the nucleus (mass number $A$) can be deduced from the following formula
\cite{Wang:2022kwg,Ma:2023tsi}:
\begin{equation}
\begin{aligned}
  n_{\rm SRC}^{\rm A} = A \times a_2(A) \times n_{\rm SRC}^{\rm d} / 2,
\end{aligned}
\label{nsrc_A}
\end{equation}
where $n_{\rm SRC}^{\rm d}$ is the number of SRC pairs in deuteron.
If $n_{\rm SRC}^{\rm d}$ is determined, then the absolute
number of 2N SRC pairs in the nucleus is obtained.
Once the absolute number of 2N SRC pairs is obtained,
the 2N SRC probability in a nucleus is easily computed
using Eq. (\ref{P2N-def}).

Currently, there are some estimations on the number
of SRC pairs in the deuteron. In our previous analysis,
$n_{\rm SRC}^{\rm d}$ is determined to be $0.021\pm 0.005$
from a correlation analysis of the nuclear mass and $a_2$ \cite{Wang:2020egq}.
By counting the high-momentum nucleons of momentum above
275 MeV/c, $n_{\rm SRC}^{\rm d}$ is estimated
to be $0.041\pm 0.008$ by CLAS collaboration \cite{CLAS:2005ola}.
Regardless of the obvious inconsistence between the
two estimations, both values of $n_{\rm SRC}^{\rm d}$
from the models are used in this analysis.
The resulting 2N SRC probabilities based on
the values of $n_{\rm SRC}^{\rm d}$ from the two models
are denoted as $P_{\rm 2N}^{\rm I}$ and $P_{\rm 2N}^{\rm II}$, respectively.
The computed 2N SRC probabilities of various nuclei are listed
in Table \ref{tab:P2N-and-nuclear-mass}.

\begin{figure}[htbp]
\begin{center}
\includegraphics[width=0.47\textwidth]{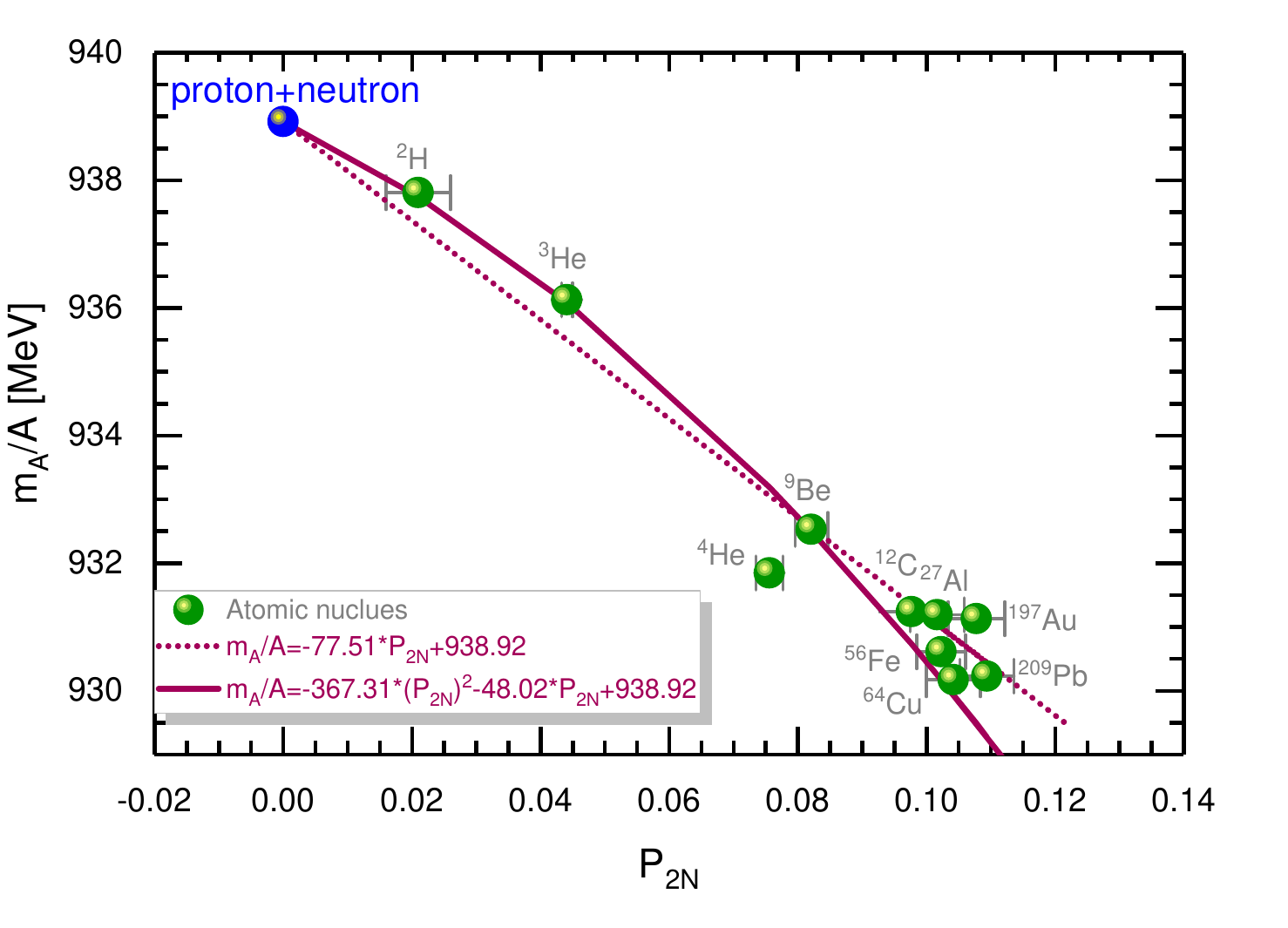}
\caption{The per-nucleon nuclear mass $m_{\rm A} / A$ as a function of
         the 2N SRC probability $P_{\rm 2N}^{\rm I}$ from our previous estimation \cite{Wang:2020egq}.
         The solid and dashed curves respectively show the fits
         with and without 3N short-range correlations. }
\label{fig:fit-mass-vs-P2N-I}
\end{center}
\end{figure}

\begin{figure}[htbp]
\begin{center}
\includegraphics[width=0.47\textwidth]{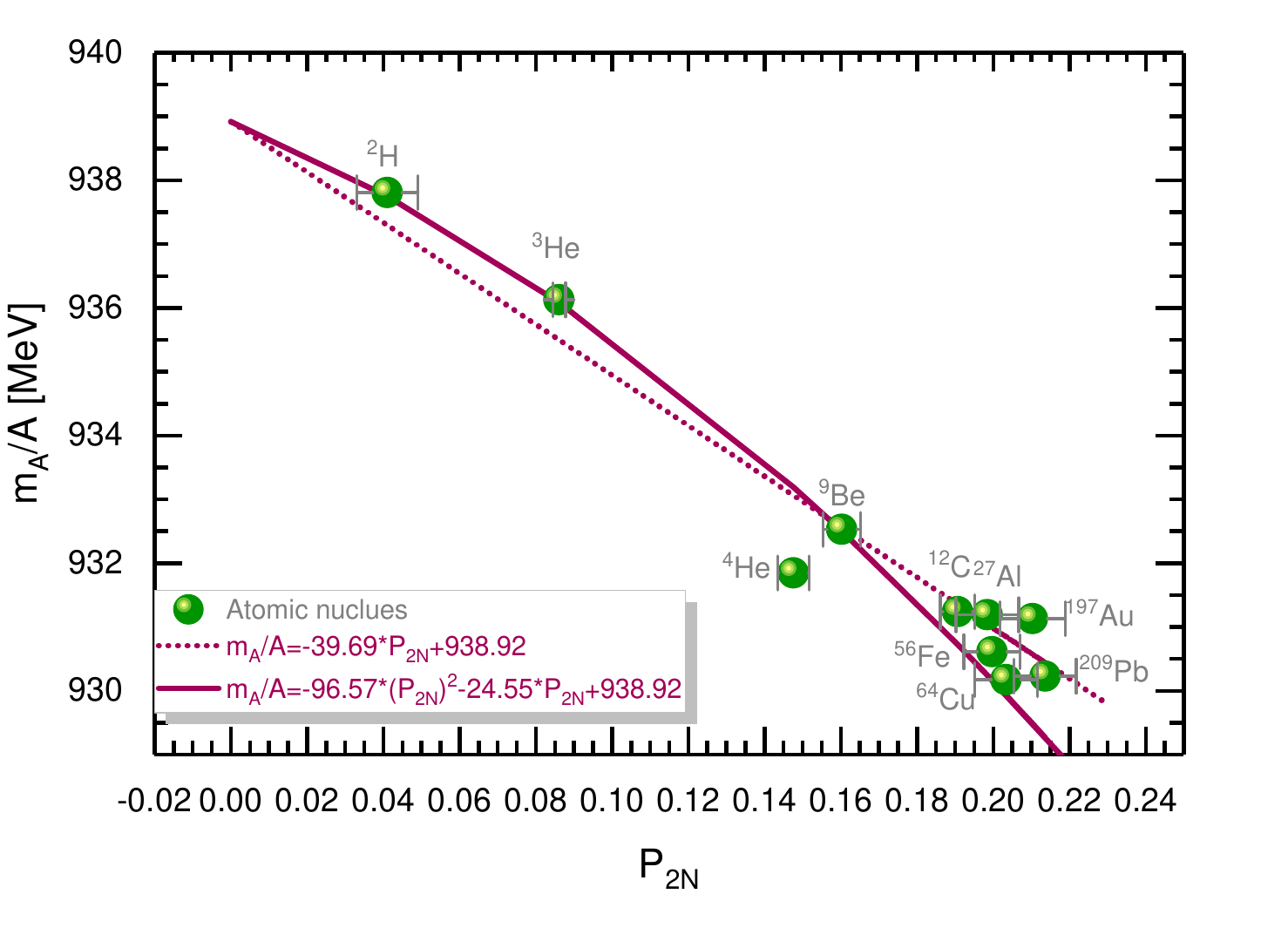}
\caption{The per-nucleon nuclear mass $m_{\rm A} / A$ as a function of
         the 2N SRC probability $P_{\rm 2N}^{\rm I}$ from CLAS' estimation \cite{CLAS:2005ola}.
         The solid and dashed curves respectively show the fits
         with and without 3N short-range correlations. }
\label{fig:fit-mass-vs-P2N-II}
\end{center}
\end{figure}

For the nuclear masses of the studied nuclei,
they are very precisely measured in experiments
and the combined average values are provided
by atomic mass evaluation group.
We take the nuclear mass data from
the up-to-date evaluation in Ref. \cite{Huang:2021nwk,Wang:2021xhn}.
The resulting per-nucleon nuclear masses of
the studied nuclei are listed in Table \ref{tab:P2N-and-nuclear-mass}.

Finally, the per-nucleon nuclear mass as a function of the 2N SRC probability
$P_{\rm 2N}^{\rm I}$ is shown in Fig. \ref{fig:fit-mass-vs-P2N-I}.
Two fits to the correlation between the nuclear mass and
$P_{\rm 2N}^{\rm I}$ are performed based on the models with
and without the 3N SRC clusters. The solid and dashed curves
respectively show the fits of the two models with and without 3N SRCs.
One sees that the model with 3N SRC fits much better the experimental
data regarding the nuclear mass and the 2N SRC probability.
The quality of the fit $\chi^2/N$ are 58.37/10 and 160.5/10
for the fits with and without 3N SRC, respectively.
The sum of residual error squares is 2.35 for the linear fit without 3N SRC
configurations, while it is 1.90 for the quadratic fit with 3N SRC configurations.
These analyses indicate that there is a positive sign
of the 3N SRC clusters in the nucleus.
From the fit with 3N SRC configuration, we also determined
the free parameters: $(m_{\rm \bar{N}}^{\rm 2N}-m_{\rm \bar{N}})=-48.0\pm 5.0$ MeV
and $\kappa(m_{\rm \bar{N}}^{\rm 3N}-m_{\rm \bar{N}})=-367\pm 68$ MeV.
The mass deficit of 2N short-range correlated nucleon is smaller
than our previous estimation \cite{Wang:2020egq}.
The mass of 3N SRC nucleon is $572\pm68$ MeV
if $\kappa = 1$, and it is $755\pm39$ MeV if $\kappa = 2$.
These obtained masses of 2N and 3N SRC nucleons would
provide some information on the modified inner structures of SRC nucleons.
As the SRC nucleon mass is much smaller than the mean-field nucleon mass,
the mean-field nucleon mass is taken as the average of free
proton and free neutron masses, $m_{\rm \bar{N}} =938.92$ MeV.
In this way, the mass decomposition formula has a fixed
point of the free nucleon ($m(P_{\rm 2N} = 0) = m_{\rm \bar{N}}$).

Fig. \ref{fig:fit-mass-vs-P2N-II} shows the per-nucleon nuclear mass
as a function of the 2N SRC probability $P_{\rm 2N}^{\rm II}$
from CLAS' estimation \cite{CLAS:2005ola}.
We also performed two fits to the correlation
between the nuclear mass and $P_{\rm 2N}^{\rm II}$ based on the models
with and without the 3N SRC clusters. With no surprise,
the model with 3N SRC fits much better the experimental
data regarding the nuclear mass and $P_{\rm 2N}^{\rm II}$.
The quality of the fit $\chi^2/N$ are 58.39/10 and 161.3/10
for the fits with and without 3N SRC, respectively.
Judged by the analysis of the correlation between nuclear mass
and $P_{\rm 2N}^{\rm II}$, there is a strong sign of 3N SRC clusters in the nucleus.
Although the obtained free parameters are different
from those obtained from the fit to $P_{\rm 2N}^{\rm I}$ data,
the conclusion on the existence of 3N SRC is the same.

The fitting results of the fits discussed above
are all summarized in Table \ref{tab:fitting-results},
including the quality of the fit, the extracted values of the free parameters.

\begin{table}[h]
\caption{The fitting results of the correlation between the nuclear mass
         and the 2N SRC probability. The free parameters in the two models
         (with and without 3N SRC) are extracted and listed here.
         See the main text for more details of the analysis. }
\renewcommand\arraystretch{1.25}
  \begin{tabular}{cccc}
    \hline\hline
      Fit setting & $\chi^2$ & $(m_{\rm \bar{N}}^{\rm 2N}-m_{\rm \bar{N}})$ & $\kappa(m_{\rm \bar{N}}^{\rm 3N}-m_{\rm \bar{N}})$ \\
    \hline
       $P_{\rm 2N}^{\rm I}$ data, & \multirow{2}{*}{58.37} & \multirow{2}{*}{$-48.0\pm5.0$ MeV} & \multirow{2}{*}{$-367\pm68$ MeV} \\
       with 3N SRC                &&& \\
    \hline
       $P_{\rm 2N}^{\rm I}$ data, & \multirow{2}{*}{160.5} & \multirow{2}{*}{$-77.5\pm1.2$ MeV} & \multirow{2}{*}{/} \\
       w/o 3N SRC                 &&& \\
    \hline
       $P_{\rm 2N}^{\rm II}$ data, & \multirow{2}{*}{58.39} & \multirow{2}{*}{$-24.6\pm1.7$ MeV} & \multirow{2}{*}{$-97\pm12$ MeV} \\
       with 3N SRC                &&& \\
    \hline
       $P_{\rm 2N}^{\rm II}$ data, & \multirow{2}{*}{161.3} & \multirow{2}{*}{$-39.7\pm0.4$ MeV} & \multirow{2}{*}{/} \\
       w/o 3N SRC                 &&& \\
    \hline\hline
  \end{tabular}
\label{tab:fitting-results}
\end{table}

\section{Discussions and summary}\label{sec:discussions-summary}

In this work, we find a positive sign of the existence of
3N SRC cluster in the nucleus, by the analysis of
the correlation between the nuclear mass and the 2N SRC probability.
A quadratic function is much better in describing the correlation
between the mass and the 2N SRC probability, compared to the linear function.
The quadratic function is explained with a nuclear mass decomposition
formula considering there are three types of nucleon in the nucleus:
the mean-field nucleon, the 2N SRC nucleon and the 3N SRC nucleon.
It is worth noting that the absolute 2N-SRC probability is derived
from the experimental measurements of $a_2$ and the number of
2N SRC pairs in the deuteron. Hence, the absolute 2N-SRC probability is
more or less the experimental determined quantity.
Actually, no matter how large the number of 2N SRC pairs in the deuteron is,
the correlation between the nuclear mass and the 2N SRC probability is
better explained with the model considering the existence of 3N SRC clusters.
Therefore, we conclude that there is a sign of 3N SRC,
showing in the the experimental data on the 2N SRC probabilities
of the studied nuclei: $^2$H, $^3$He, $^4$He, $^9$Be,
$^{12}$C, $^{27}$Al, $^{56}$Fe, Cu, $^{197}$Au and $^{208}$Pb.

The analysis demonstrated in this work should be treated
as a preliminary study. The nuclear mass decomposition formula
used in the correlation analysis of the mass and the 2N SRC probability
is just an approximate equation.
The formula is based on three main assumptions:
(i) there are 2N and 3N SRCs in the nucleus in addition to
the independent mean-field nucleons;
(ii) the masses of 2N SRC and 3N SRC nucleons are
universal quantities in different nuclei;
(iii) the 3N SRC is formed from the cohesion
of a nucleon and a 2N SRC pair.
Therefore, the analysis result present in this work is model-dependent.

From the analysis, the masses of 2N SRC and 3N SRC nucleons are
also extracted from the fits.
We find that the extracted SRC nucleon mass is very sensitive
to the absolute probability of 2N SRC applied in the analysis.
Nevertheless, the SRC nucleon mass extracted in this work
would provide some valuable references on
the nuclear modification of the inner structure of SRC nucleon.

At the end, we want to point out that the nonlinear correlation
between the nuclear mass and the 2N SRC probability
may also arise from 4N SRC, or $\alpha$ cluster,
or even much bigger cluster of more than four nucleons.
If the multi-nucleon cluster is formed from
the sequential processes of nucleon cohesion,
then the bigger cluster has the smaller formation probability.
Hence, the the nonlinear correlation between the mass
and the 2N SRC probability is mainly attributed to the 3N SRC cluster.
There is still some rooms for an improvement of the current analysis.
We can imagine that the nucleus is much complex quantum
system of many nucleons, and there are much diversified and unknown
micro-structures inside it.
We still need more theoretical developments in revealing
the underlying mechanism of 3N SRC, and the novel experimental
techniques to confirm the existence of 3N SRC,
and even to study the properties of these minorities in the nucleus.

\begin{acknowledgments}
N.-N. Ma is supported by the National Natural Science Foundation of China under the Grant NO. 12105128.
R. Wang is supported by the National Natural Science Foundation of China under the Grant NO. 12005266
and the Strategic Priority Research Program of Chinese Academy of Sciences under the Grant NO. XDB34030301.
\end{acknowledgments}

\bibliographystyle{apsrev4-1}
\bibliography{refs}

\end{document}